\documentclass[APS, twocolumn, a4paper, amsfonts, showpacs]{revtex4}
\usepackage{dcolumn,amsmath,xspace}
\usepackage{graphicx}
\usepackage{bm}
\usepackage{color}

\begin{document}
\title{Integer Quantum Hall Effect in Trilayer Graphene}

\author{A. Kumar$^1$, W. Escoffier$^1$\footnote{Author to whom correspondance should be addressed}, J.M. Poumirol$^1$, C. Faugeras$^2$, D. P. Arovas$^3$, M. M. Fogler$^3$, F. Guinea$^4$, S. Roche$^{5,6}$, M. Goiran$^1$ and B. Raquet$^1$}

\affiliation{$^1$ Laboratoire National des Champs Magn\'etiques Intenses (LNCMI), CNRS-UPR3228, INSA, UJF, UPS, Universit\'e de Toulouse, 143 av. de rangueil, 31400 Toulouse, France}%
\affiliation{$^2$ Laboratoire National des Champs Magn\'etiques Intenses (LNCMI), CNRS-UPR3228, INSA, UJF, UPS, 25 rue des Martyrs, 38042 Grenoble, France}%
\affiliation{$^3$ Department of Physics, University of California at San Diego, 9500 Gilman Drive, La Jolla, California 92093, USA}%
\affiliation{$^4$ Instituto de Cienca de Materiales de Madrid, CSIC, Cantoblanco E28049 Madrid, Spain}%
\affiliation{$^5$ CIN2 (ICN-CSIC) and Universidad Autonoma de Barcelona, Catalan Institute of Nanotechnology, Campus UAB, 08193 Bellaterra (Barcelona), Spain}%
\affiliation{$^6$ ICREA, Instituci$\acute{o}$ Catalana de Recerca i Estudis Avan\c cats, Barcelona, Spain}%
\date{\today}

\begin{abstract}
The Integer Quantum Hall Effect (IQHE) is a distinctive phase of two-dimensional electronic systems subjected to a perpendicular magnetic field.  Thus far, the IQHE has been observed in semiconductor heterostructures and in mono- and bi-layer graphene.  Here we report on the IQHE in a new system: trilayer graphene.  Experimental data are compared with self-consistent Hartree calculations of the Landau levels for the gated trilayer.  The plateau structure in the Hall resistivity determines the stacking order (ABA {\it versus\/} ABC).  We find that the IQHE in ABC trilayer graphene is similar to that in the monolayer, except for the absence of a plateau at filling factor $\nu=2$. At very low filling factor, the Hall resistance vanishes due to the presence of mixed electron and hole carriers induced by disorder.

\end{abstract}


\maketitle

\begin{figure*}[ht!]
\includegraphics[width=2\columnwidth]{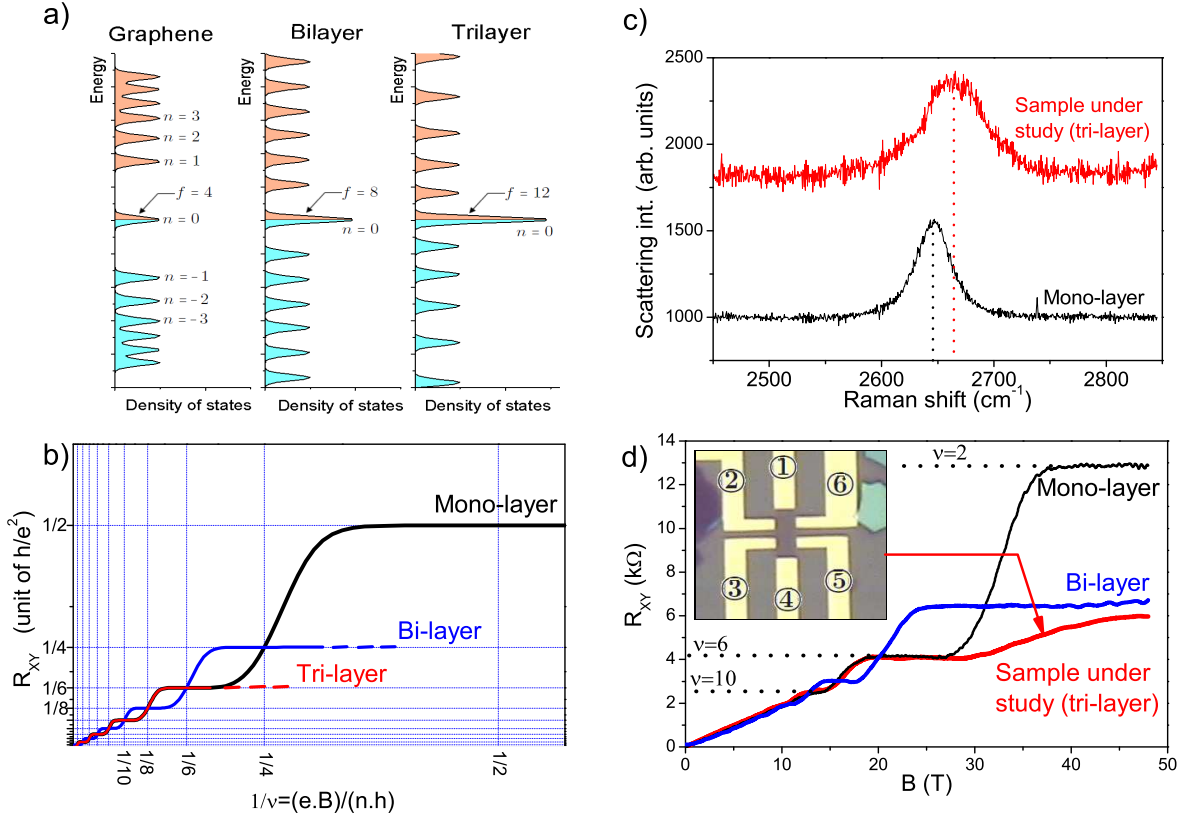}
\caption{Identifying the sample under study as a trilayer graphene.  a) and b) schematic representations of the IQHE in mono-, bi- and tri-layer graphene. The degeneracy of the zeroth Landau Level, equally shared by electrons and holes,
is $f=g$-fold degenerate in graphene, $f=2g$-fold degenerate in bilayer graphene and $f=3g$-fold degenerate in trilayer
graphene where $g=4$, leading to a sequence of the Hall resistance plateaus successively shifted by $2h/e^2\Omega$.
Possible LL degeneracy lifting at high field is not represented in these drawings.  c) Raman spectrum of trilayer graphene
and mono-mayer graphene measured on the same substrate. d) Experimental IQHE in tri, bi and mono-layer graphene
for equivalent carrier density $n=3.4\times 10^{12}\,{\rm cm}^{-2}$ and quasi-equivalent mobility. The optical image of the
trilayer graphene sample is shown in insert. As contact 1 was proven defective, a constant current of $i=1\,\mu{\rm A}$ is
injected through contacts 2 and 4. The Hall resistance is measured between contacts 3 and 5.}
\label{fig.1}
\end{figure*}

More than 30 years after its initial discovery in semiconductor two dimensional electron gases (2DEG), the Integer Quantum Hall Effect (IQHE) remains one of the most fascinating phenomena in condensed matter physics.  The recent discovery of graphene \cite{Novoselov26072005} boosted this research field by providing a new 2D system where Dirac-like electronic excitations with Berry's phase $\pi$ leads to a new form of IQHE \cite{Nature.438.197.2005, Nature.438.201.2005}, with plateaus at $\sigma_{xy}=(n+\frac{1}{2})\,ge^2/h$, where $g$ is the Landau level degeneracy due to spin and valley degrees of freedom.  Soon afterward, a third type of IQHE was reported in bilayer graphene, where the $2\pi$ Berry's phase of charge carriers results in a conventional quantization sequence, except that the last Hall plateau is missing \cite{NaturePhysics.2.177.2006}. As the dynamics of charged carriers change every time an extra graphene layer in added, it was theoretically anticipated that the Landau Level (LL) spectrum of $N$-layer graphene systems would result in distinctive IQHE features arising from an $N\pi$ Berry's phase.  In trilayer graphene, the zero-energy LL is expected to be 12-fold degenerate so that the Hall resistance plateau sequence follows the same ladder as in graphene, but the plateau at $\nu=\pm 2$ should be missing (see figures \ref{fig.1}-a and \ref{fig.1}-b).  So far, most of the studies dedicated to IQHE in trilayer graphene have been restricted to theoretical calculations \cite{PhysRevB.77.155416, JPSJ.76.094701, PhysRevB.77.045429, PhysRevB.78.033403, Aoki2007123}. Indeed, experimental realizations are difficult since the knowledge of the exact number of layers and their relative stacking order are challenging to ascertain.  Making use of both high field magneto-transport and Raman spectroscopy, we clearly identified a trilayer graphene sample and report on a fourth type of IQHE in this system. Self-consistent Hartree calculations of the gated trilayer Landau levels based on the Slonczewski-Weiss-McClure (SWMC) tight binding model \cite{PhysRev.109.272, PhysRev.108.612, Carbon.7.425.1969} have been performed and favorably compared to the experimental data.  Thus, a comprehensive knowledge of the underlying electronic properties of trilayer graphene is presented, together with an unambiguous determination of the stacking order.

Many few layer graphene flakes were deposited onto a $d=280\,$nm thick thermally grown silicon oxide on silicon substrate using the standard micro-mechanical exfoliation of natural graphite.  The flake measured in this study was first
roughly identified as few-layer graphene using optical microscopy.  The Raman scattering spectrum was measured at
room temperature using a confocal micro Raman scattering set-up using He-Ne laser excitation ($\lambda=632.8\,$nm)
with $\sim 1\,$mW optical power focused on a 1 $\mu$m diameter spot.  The 2D band feature (also called ${\rm G}'$
feature) of this sample is shown in figure \ref{fig.1}-c and appears in the form of a multicomponent feature characteristic
of multi-layer graphene specimens \cite{Graf200744}, different from the one observed for a mono-layer graphene
specimen processed in the same way (also shown in figure \ref{fig.1}-c). The experimental IQHE of the sample under
study is displayed in figure \ref{fig.1}-d, together with bi-layer and mono-layer graphene fingerprints of other samples.
These samples have an equivalent carrier density and similar mobility (see legend of figure \ref{fig.1} for details).  For the
sample under consideration, the sequence of the Hall resistance plateaus is described by $R_{xy}=h/\nu e^2$ where $\nu=6$,
$10$, $14...$. Taking into account the results of both Raman spectroscopy and IQHE, we thus unambiguously identify the sample as trilayer graphene.  This assertion is further reinforced when comparing the optical contrast of the flake with other mono-layer graphene flakes under microscope
(see the supplementary information). As expected, the IQHE in trilayer graphene is indistinguishable from its monolayer
counterparts except at very high field where the $\nu=2$ quantized Hall resistance plateau is absent.

\begin{figure}[ht!]
\includegraphics[width=\columnwidth]{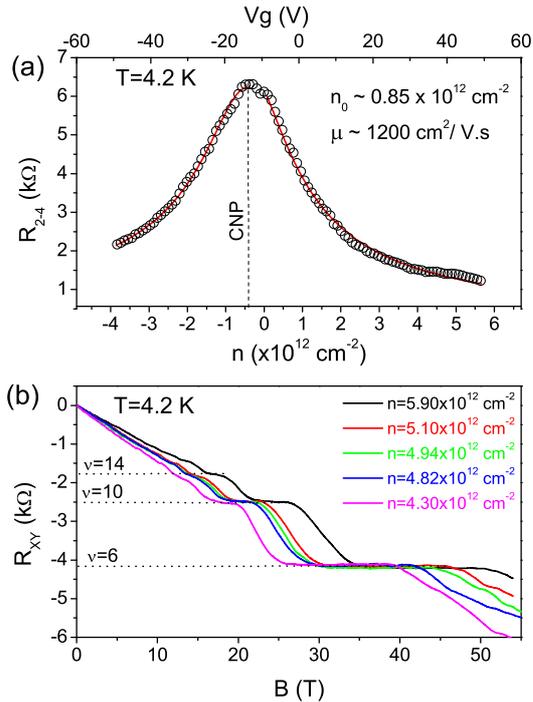}
\caption{a) Electrical resistance of the sample as a function of carrier concentration at $T=4.2\,$K. Theoretical
adjustment of the experimental data using \textbf{\cite{kim:062107}} together with the deduced field-effect mobilities at residual carrier density of $0.85 \times 10^{12}\, {\rm cm}^{-2}$ b) Quantized Hall resistance profile in trilayer graphene
for selected charge carrier density far from CNP.}
\label{fig.2}
\end{figure}

Before further proceeding with a detailed presentation of IQHE in graphene trilayer, let us review some important
experimental considerations. Once identified, the sample was connected to multi-terminal electrodes made by successive
thermal evaporation of Ti [5 nm] and Au [40 nm] through a PMMA mask (see figure \ref{fig.1}-d, inset). The flake was
further patterned in the Hall bar geometry using electron beam lithography and oxygen plasma etching. The contact
resistances were estimated to a few hundred ohms each. A gate voltage $V_{\rm g}$ between the sample and the
substrate was used to electrically induce electrons or holes up to densities of $6\times 10^{12}\,{\rm cm}^{-2}$.
The sample was annealed alternatively in vacuum and in helium gas at a temperature T=110$^\circ\,$C. The induced carrier
density as a function of the gate voltage follows the simple plane capacitor model $n=\kappa\,\epsilon_0
(V_{\rm g}-V_{\rm CNP})/ed$ where $\epsilon_0$ is the vacuum dielectric permittivity, $\kappa=3.9$ is the
relative permittivity for SiO$_2$ and $V_{\rm CNP}=-13.75\,$V is the gate voltage required to reach the Charge Neutrality
Point (CNP), indicating the presence of a $n$-type residual doping estimated to $n_0=0.85 \times 10^{12}\,{\rm cm}^{-2}$.
Figure \ref{fig.2}-a shows the typical two-probe longitudinal resistance as a function of the carrier density at $T=4.2\,$K,
while figure \ref{fig.2}-b displays the Hall resistance for selected carrier concentrations far from CNP. The lower bound (Hall) mobility is estimated to $\mu_{\rm e} \sim 1300\pm100\,{\rm cm}^2.{\rm V^{-1}}\!\cdot\! {\rm s^{-1}}$ for electrons and $\mu_{\rm h} \sim 900\pm300\, {\rm cm}^2.{\rm V^{-1}}\!\cdot\!{\rm s^{-1}}$ for holes. The Hall resistance displays well defined
plateaus at $R_{xy}=h/(\nu e^2)$ with $\nu=6$, $10$, and $14$. We notice that the Hall resistance slightly overshoots
the $\nu=6$ resistance plateau at very high magnetic field. This surprising feature triggered the need for a
detailed theoretical analysis of the LL spectrum \cite{PhysRevB.73.245426, PhysRevB.78.245416} which goes beyond the simple model
presented earlier in the introduction.\\

\begin{figure*}
\begin{minipage}[c]{0.6\linewidth}
\includegraphics[width=\linewidth]{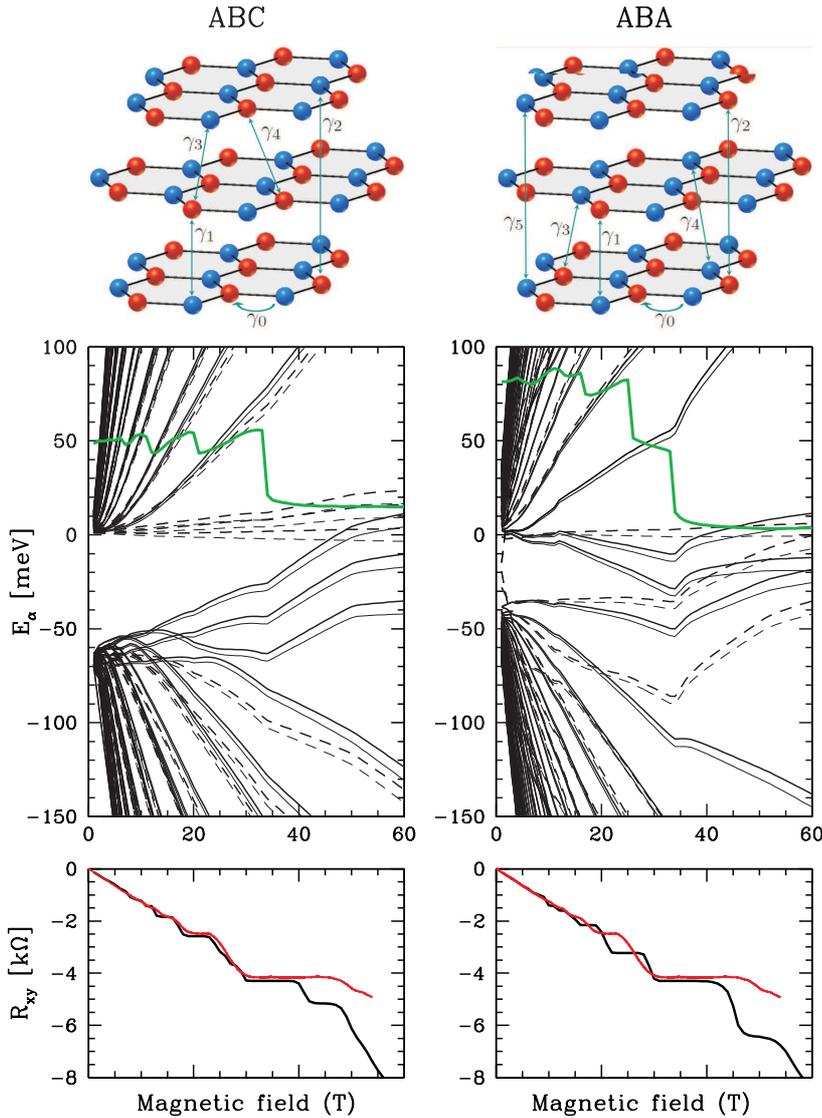}
\end{minipage}\hfill
\begin{minipage}[c]{0.4\linewidth}
\caption{a) Illustration of ABC-stacked trilayer graphene (top panel), theoretical Landau level structure (middle panel) and quantized Hall resistance (bottom panel) using the following parameters : $V_{\rm g}=50$ V, $V_{\rm CNP}=-13.75$ V, $T=4.2$ K, $W_0=10$ meV and $g=2$. In the middle panel, the solid and dashed curves indicate the Landau levels originating from valleys K and K', respectively. In the bottom panel, the experimental Hall resistance is displayed as the red curve, superimposed with theoretical results as the black line. b) For comparison, the same information for the ABA stacked trilayer graphene}
\label{fig.3}
\end{minipage}
\end{figure*}

To model the gated trilayer, we employ an SWMC tight-binding parametrization of the local hopping amplitudes \cite{PhysRev.109.272, PhysRev.108.612} and treat the Coulomb interactions via a self-consistent Hartree approach. There are two possible stacking orders to consider: Bernal (ABA) and rhombohedral (ABC), both illustrated in figure \ref{fig.3}.  For the Bernal case, we take $\gamma_0=3000\,$meV, $\gamma_1=400\,$meV, $\gamma_2=-20\,$meV,
$\gamma_3=300\,$meV, $\gamma_4=150\,$meV, and $\gamma_5=38\,$meV.  In addition, there is an on-site energy shift of $\frac{1}{2}\Delta=18\,$meV for each $c$-axis neighbor.  For the rhombohedral case \cite{Carbon.7.425.1969}, the parameter $\gamma_5$ does not enter. When a gate voltage is applied to the device, the charge carriers will be distributed among the three layers. Integrating Gauss' law across the layers provides a set of three equations which have to be solved self-consistently in order to obtain the trilayer potentials (see supplementary informations for details).  We assume that the system contains stray charges which are located on the top layer of the device. These charges are responsible for the offset voltage $V_{\rm CNP}$ necessary to bring the system to neutrality.  The zero temperature Hall conductivity shifts by $e^2/h$ when the electrochemical potential crosses the center of a Landau level.  We include Zeeman splitting, assuming $g=2$. Finally, some disorder is introduced in the model through a convolution of the density of states with a square distribution of half-width $W_0=10\,$meV, estimated from the sample's mobility. 

Figure \ref{fig.3} shows the LL energies and theoretical Hall resistance for both ABC and ABA stacking for gate voltage $V_{\rm g}=+50\,$V, along with the experimental results for $R_{xy}$ (the full set of IQHE data is reported in the supplementary information).  For fields up to $40\,$T, the measurements agree fairly well with the theoretical predictions for the ABC trilayer, and fail to reproduce the theoretical Hall plateau sequence for the ABA trilayer. The contrasting plateau sequences for ABC and ABA trilayers arise due to the significant differences in their respective Landau level
structures, as is evident in figure \ref{fig.3}. Indeed, the rhombohedral stacking order accounts for the absence of Hall plateaus at some filling factors, like $\nu=8$ and $\nu=12$. A similar effect occurs in monolayer graphene, where the plateaus at $\nu=4$, $8$, $12$, $\ldots$ are missing due to valley degeneracy arising from the inversion symmetry of the honeycomb lattice.  With no bias voltage, the ABC trilayer is inversion symmetric, while the ABA trilayer is not \cite{PhysRevB.81.115315}. The presence of an electric field across the graphene layers, due to the gate voltage induced charge redistributions \cite{Aoki2007123, PhysRevB.81.125304}, breaks the lattice inversion symmetry. However, this field-induced splitting between LLs arising from different valleys is small in the ABC case, except for the six levels closest to the Dirac energy. Neglecting Zeeman splitting, quantum Hall steps of amplitude $\Delta \sigma_{xy}=2.e^2/h$ should be observed for each plateau-to-plateau transition. This holds in particular for the Bernal type of stacking as the LLs originating from valleys $K$ and $K'$ are quite distinct from each other due to the absence of inversion symmetry.  On the other hand, the ABC-stacked LL band-structure is much less affected by electrostatic effects. In high enough magnetic field, the LLs evolve roughly by bunches of four and, when disorder effects are taken into account, lead to quantum Hall steps of $\Delta \sigma_{xy}=4.e^2/h$, as experimentally observed. It has been demonstrated that the stacking in graphene trilayers can be analyzed making use of micro- Raman spectroscopy \cite{doi:10.1021/nl1032827}. However, the Raman spectral signature of stacking is quite subtle and undoubtedly inconclusive for the present experiment (figure \ref{fig.1}-c). Although it has been found that ABC-stacked graphene trilayer flakes are much rarer than their ABA-stacked counterpart using the micro-mechanical cleavage method, the theoretical analysis of IQHE allows an unambiguous determination of the rhombohedral staking order for the sample considered here.

\begin{figure}
\includegraphics[width=\columnwidth, bb=75 15 290 260]{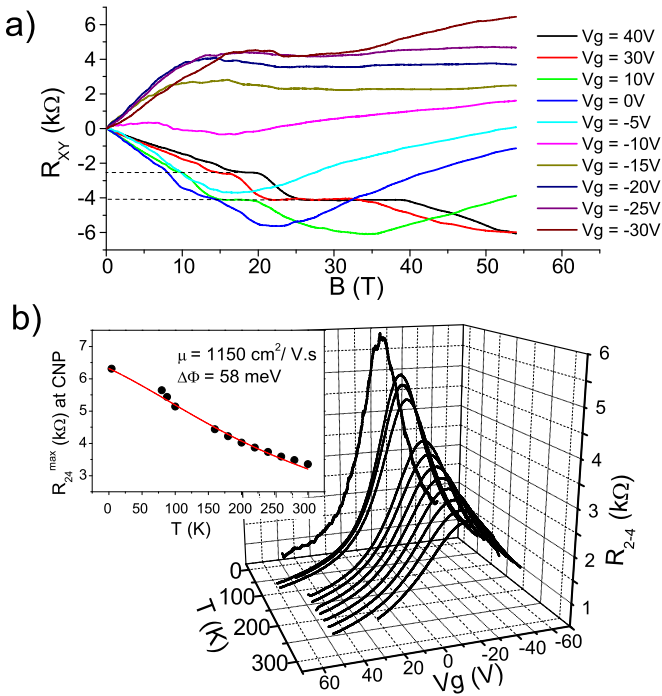}
\caption{a) Hall resistance for various back gate voltage in the close vicinity of the CNP. Notice that the Hall resistance tends to vanish at very high magnetic field as the ratio between electrons and holes tends to equilibrate. b) Longitudinal resistance $R_{2-4}$ as a function of gate voltage and temperature. The maximum resistance as a function of temperature is shown in the insert. A theoretical fit according to the model developed in \cite{PhysRevB.82.081409} is also shown. The theoretical adjustment takes into account the increase of carriers density with temperature and the resulting activated conductance through an inhomogeneous system made of electron/hole puddles.}
\label{fig.4}
\end{figure}

Interestingly, the IQHE fails to be reproduced at very low filling factor (high magnetic field and low carrier density). To further investigate this issue, we analyze the Hall resistance for various charge carrier concentrations close to CNP. Focusing at figure \ref{fig.4}-a, we begin the analysis with the Hall resistance for $V_{\rm g}=+40\,$V ($n=4.3\times 10^{12}\,{\rm cm}^{-2}$), which displays well defined quantized Hall plateaus.  As the gate voltage is decreased, the corresponding set of curves is shifted so that the quantized plateaus occur at lower magnetic field, as expected from IQHE theory.  On the other hand, as the Fermi energy is driven closer to the CNP, the low field Hall effect is no longer linear reflecting the presence of electrons and holes that both contribute to transport.  In the range $-50\,{\rm V} < V_{\rm g} < +20\,$V, the initial ratio between electron/hole density evolves as the magnetic field is increased to accommodate the field-induced redistribution of quantum states available in the lowest LL \cite{PhysRevB.82.121401}. Actually, the electron and hole densities tend to equilibrate and consequently the Hall resistance vanishes at high field. This effect is a hallmark of the disordered 2DEG \cite{PhysRevLett.104.166401}, where the presence of electron and hole puddles allows both types of carriers for a given Fermi energy close to CNP.  Indeed, the inevitable presence of charged impurities or lattice imperfections introduces long-range disorder and leads to a spatially inhomogeneous and fluctuating potential landscape, resulting in some local accumulation of charge carriers \cite{NatPhys.4.144.2008, NatPhys.6.74.2010, PhysRevB.76.195421, PhysRevLett.103.236801, PhysRevB.79.245423}. Alternatively, its consequences on charge transport can be monitored through temperature-dependent measurements.  The main frame of figure \ref{fig.4}-b shows the two-probe longitudinal resistance $R_{2-4}$ with varying gate voltage and temperature.  The resistance increases with decreasing temperature in the vicinity of the CNP while it remains constant for higher doping. This trend is consistent with what has been previously observed in disordered-like samples, for which the highly inhomogeneous carrier density can lead to both metallic and activated transport. The temperature evolution of the resistance maximum at CNP is given in the inset of figure \ref{fig.4}-b.  A good agreement with the predictions of reference \cite{PhysRevB.82.081409} is achieved with fitting parameters $\mu=1150\,{\rm cm}^2/{\rm V}\!\cdot\!{\rm s}$ and $\Delta\Phi=58\,$meV, where $\Delta\Phi$ stands for the rms amplitude of the fluctuating potential landscape.  This value is slightly larger than those reported in the literature for bilayer graphene \cite{PhysRevB.82.081407, AppPhysLett.95.24.2009} ($\Delta \Phi=21.5\,$meV for $\mu\approx 2000\,{\rm cm}^2/{\rm V}\!\cdot\!{\rm s}$), as predicted for FLG \cite{NatNano.6.383.2009}.\\

To summarize, we report for the first time the observation of a new form of IQHE in a gated graphene trilayer. The filling factor sequence associated with the quantized Hall resistance plateaus is identical to that for graphene, but the plateau at $\nu=2$ is missing. The experimental data are supported by a theoretical analysis where both Bernal and rhombohedral stacking have been considered. The main experimental IQHE features are reproduced only for the rhombohedral case, emphasizing the importance of stacking order in the electronic properties of graphene trilayers. Usually, graphene trilayers deposited on SiO$_2$ show a poor mobility of the order of $1000\,{\rm cm}^2/{\rm V}\!\cdot\!{\rm s}$ \cite{PhysRevB.80.235402, NatNano.6.383.2009, AppPhysLett.95.24.2009}, justifying the need for very high magnetic field for IQHE studies. Important progresses are expected in samples with higher mobility, {\it e.g.\/} in suspended or boron nitride-deposited graphene trilayers.\\

\textbf{This research was supported by EuroMagNET II program under EU Contract No. 228043, by the French National Agency for Research (ANR) under Contract No. ANR-08-JCJC-0034-01, and by the technological platform of LAAS-CNRS, member of the RTB network. One of the authors (D.P.A.) was supported by NSF grant DMR-1007028. F.G. is supported by MICINN (Spain), Grants FIS2008-00124 and CONSOLIDER CSD2007-00010. The authors are very grateful to M. Potemski for very helpful and fruitful discussions.}

\end{document}